# Pressure-induced Superconductivity in Thermoelectric Semiconductor $Mg_3Sb_2$

Cuiying Pei[1#], Yasong Wu[2#], Airan Li[3#], Juefei Wu[1#], Qi Wang[1,4], Yifan Zhu[2], Yi Zhao[1], Lingling Gao[1], Changhua Li[1], Weizheng Cao[1], Shihao Zhu[1], Mingxin Zhang[1], Yulin Chen[1,4,5], Chenguang Fu[3*], Tiejun Zhu[3], Jiong Yang[2,6*] and Yanpeng Qi[1,4,7*]

1. State Key Laboratory of Quantum Functional Materials, School of Physical Science and Technology, ShanghaiTech University, Shanghai 201210, China
2. Materials Genome Institute, Shanghai University, Shanghai 200444, China
3. State Key Laboratory of Silicon and Advanced Semiconductor Materials, School of Materials Science and Engineering, Zhejiang University, Hangzhou 310058, China
4. ShanghaiTech Laboratory for Topological Physics, ShanghaiTech University, Shanghai 201210, China
5. Department of Physics, Clarendon Laboratory, University of Oxford, Parks Road, Oxford OX1 3PU, UK
6. Zhejiang Laboratory, Hangzhou, Zhejiang 311100, China
7. Shanghai Key Laboratory of High-resolution Electron Microscopy, ShanghaiTech University, Shanghai 201210, China

# These authors contributed to this work equally.

## ABSTRACT

The intrinsic electronic structures of narrow bandgap thermoelectric (TE) materials serve as a platform for the investigation of coupling effects of quasi-particles under high pressure, enabling the exploration of emerging electronic and phonon transport, superconductivity, and topological transition. Here, we report the discovery of pressure-induced superconductivity in the TE semiconductor $Mg_3Sb_2$. Upon the increased pressure, the metallization occurs at ~8.7 GPa, followed by a superconducting transition concomitant with a carrier-type crossover from *p*- to *n*-type. This phenomenon arises from a pressure-induced structural phase transition from the semiconducting $P\overline{3}m1$ to the metallic $C2/m$-I phase. The superconducting critical temperature ($T_c$) exhibits a dome-shaped pressure dependence, peaking at 3.3 K at 12.6 GPa. Combined theoretical calculations, high-pressure Raman spectroscopy, and X-ray diffraction (XRD) measurements reveal an additional structural transition above ~20 GPa, yielding a distinct $C2/m$-II phase. Our findings establish the high-pressure phase diagram of $Mg_3Sb_2$, elucidate its pressure-dependent electronic properties, and provide valuable insights for future investigations of TE materials under high pressure.

## INTRODUCTION

Thermoelectric (TE) materials, enabling direct heat-to-electricity conversion, have emerged as promising solutions for power generation and solid-state cooling.[1-4] TE materials are generally narrow bandgap semiconductors whose electronic structure, for instance, band gap, band anisotropy, band degeneracy, and electronic density of states (DOS), can be tunable through isoelectronic alloying, aliovalent doping, or temperature changing. These offer typical band engineering strategies that have significantly contributed to the enhancement of TE performance in the past two decades.[5-7] In addition, pressure acts as a clean and effective tool to tune the electronic structure and electron-phonon coupling of solid materials and enables the induction of novel quantum phenomena.[8-12] This makes high-pressure studies on narrow bandgap semiconductor TE materials ideal for exploring comprehensive couplings among lattices, electrons, and spins, providing a route to potentially enhance the TE properties, as well as inducing emerging physical phenomena, such as superconductivity and topological phase transition.[13-16]

Among TE systems, $Mg_3Sb_2$-based compounds stand out for their high TE performance, low cost, and good mechanical properties.[17-23] As one member of the $AB_2X_2$ family, $A$ and $B$ sites in $Mg_3Sb_2$ are occupied by Mg atoms, while $X$ corresponds to the Sb atom. These Mg atoms reside in distinct sublattices, denoted as octahedral $Mg_1$ and tetrahedral $Mg_2$, respectively[24]. These compounds feature weak $Mg_1$-Sb bonds that enhance anharmonicity and reduce lattice thermal conductivity $\kappa_L$[25,26]. In addition, they possess a unique electronic structure with six degenerate conduction bands[18,27]. These traits drive extensive efforts to optimize their TE performance *via* band engineering[28-31], grain boundary modification[28,32,33], and carrier density tuning[34-36], fostering advances in both the deepened understanding of atomic occupation in $Mg_3Sb_2$[37] and the development of $Mg_3Sb_2$-based TE devices[38-42].

Under high pressure, $Mg_3Sb_2$ undergoes a reversible structural phase transition involving significantly different bond strengths of $Mg_1$-Sb and $Mg_2$-Sb.[43,44] However, the transport properties under high pressure remain underexplored despite their key role in understanding materials' electronic structures and the structure-property relationships under high pressure. Herein, we combine *in situ* high-pressure transport,

Raman and XRD measurements with structure predictions along with first-principles calculations to address this gap. We observe a semiconductor-to-metal transition at around 8.7 GPa, followed by structural transition-induced superconductivity in $Mg_3Sb_2$. The relation between pressure and superconducting transition temperature $T_c$ is dome-shaped with a maximum value of 3.3 K at 12.6 GPa. Hall effect measurements suggest that the superconductivity transition is accompanied by a carrier-type transition from *p*-type to *n*-type. With increasing pressure, Raman and XRD experiments verify signals of another structural phase transition around 20 GPa. Combining machine learning graph theory accelerated crystal structure search and first-principles calculations, we identify a new high-pressure *C*2/*m* phase. Our study is helpful to understand the phase diagram of $Mg_3Sb_2$ under high pressure, shedding light on the research on TE materials under high pressure.

## RESULTS AND DISCUSSION

The as-synthesized $Mg_3Sb_2$ single crystal displays semiconducting behavior at ambient pressure. (**Figures S1-S2**) Temperature-dependent resistivity ($\rho(T)$) under various pressures is shown in **Figure 1a**. As pressure increases, the material gradually changes from a semiconductor to a metal, accompanied by a sudden drop in $\rho$ around 1.8 K at 8.7 GPa (**Figure 1b**). Further compression leads to zero resistivity above 10.6 GPa, indicating the character of superconductivity. The $T_c$ increases with pressure, reaching a maximum of 3.3 K at ~12.6 GPa, and then gradually decreases to form a typical dome-like relationship with pressure. Measurements on different $Mg_3Sb_2$ samples in independent runs yield consistent and reproducible results, validating the metallization and superconductivity transition under high pressure (**Figures S3 and S4**).

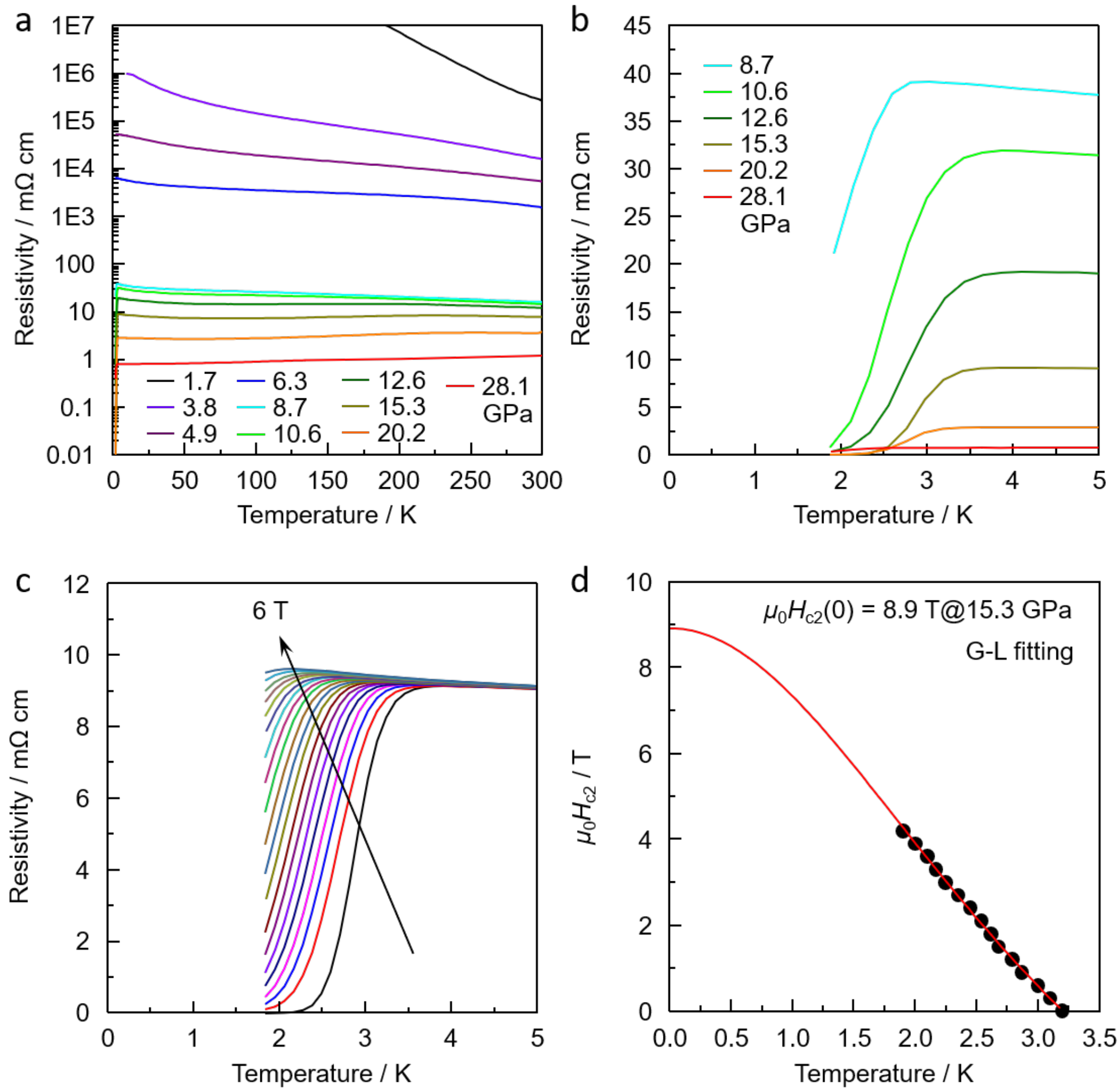

**Figure 1** (a) Temperature-dependent resistivity of $Mg_3Sb_2$ under various pressures in run I; (b) Enlarged $\rho(T)$ curve in the vicinity of the $T_c$; (c) Temperature-dependent resistivity under various magnetic fields at 15.3 GPa; (d) Temperature-dependent upper critical field of $Mg_3Sb_2$ at 15.3 GPa. Solid lines represent the fits based on the Ginzburg-Landau (G-L) formula.

We also measured the temperature-dependent resistivity of $Mg_3Sb_2$ under different magnetic fields at 15.3 GPa (**Figure 1c**). With increasing magnetic field, $T_c$ shifts toward lower temperatures, demonstrating the suppression of superconductivity by magnetic fields. We then extracted the temperature-dependent upper critical field at 15.3 GPa (**Figure 1d**) and fitted the data using the Ginzburg-Landau (G-L) formula $H_{c2}(T) = H_{c2}(0)\times(1-t^2)/(1+t^2)$, where $t=T/T_c$.[45,46] The estimated upper critical field $\mu_0 H_{c2}(0)$ is 8.9 T at 15.3 GPa, which exceeds the Pauli limit field of 6.1 T given by $H_P(0) = 1.84T_c$. According to the relationship of $H_{c2} = \Phi_0/(2\pi\xi^2)$, where $\Phi_0 = 2.07\times10^{-15}$ Wb is the flux quantum, the Ginzburg-Landau coherence length $\xi_{GL}(0)$ is derived to be 6.1 nm. Additionally, G-L fitting gives $\mu_0H_{c2}(0)$ = 8.7 T and $\xi_{GL}(0)$ = 6.2 nm at 12.6 GPa, confirming the high upper critical field in $Mg_3Sb_2$ (**Figure S4**). This observation of high $\mu_0H_{c2}(0)$ in $Mg_3Sb_2$ could be attributed to the pair-breaking effect.[47,48] Notably, high pressure can induce superconductivity in Sb with a maximum $T_c$ of 3.5 K at 8.2 GPa. However, the $\mu_0H_{c2}(0)$ of Sb is merely 0.14 T. All the above results demonstrate that the pressure-induced superconductivity in $Mg_3Sb_2$ is an intrinsic property of this material. (**Figures S5 and S6**)

To clarify the electrical properties of $Mg_3Sb_2$ under high pressure, we conducted Hall effect measurements at room temperature. Selected magnetic field-dependent Hall resistance data are shown in **Figure 2**. At 0.6 GPa, the Hall resistance curves display a linear feature with a positive slope, indicating a hole-dominated carrier type consistent with ambient-pressure measurements (**Figure 2a**). With pressure increasing, the slope of the Hall resistance curve decreases and becomes negative at about 6.7 GPa, signifying a carrier-type inversion from *p*-type to *n*-type (**Figure 2b**). The carrier-type inversion pressure (~ 6.7 GPa) aligns with the semiconductor-to-metal transition, suggesting a change in the electronic structure under high pressure. Furthermore, the absolute value of the slope decreases by several orders of magnitude above 6.7 GPa, indicating the improved electron density.

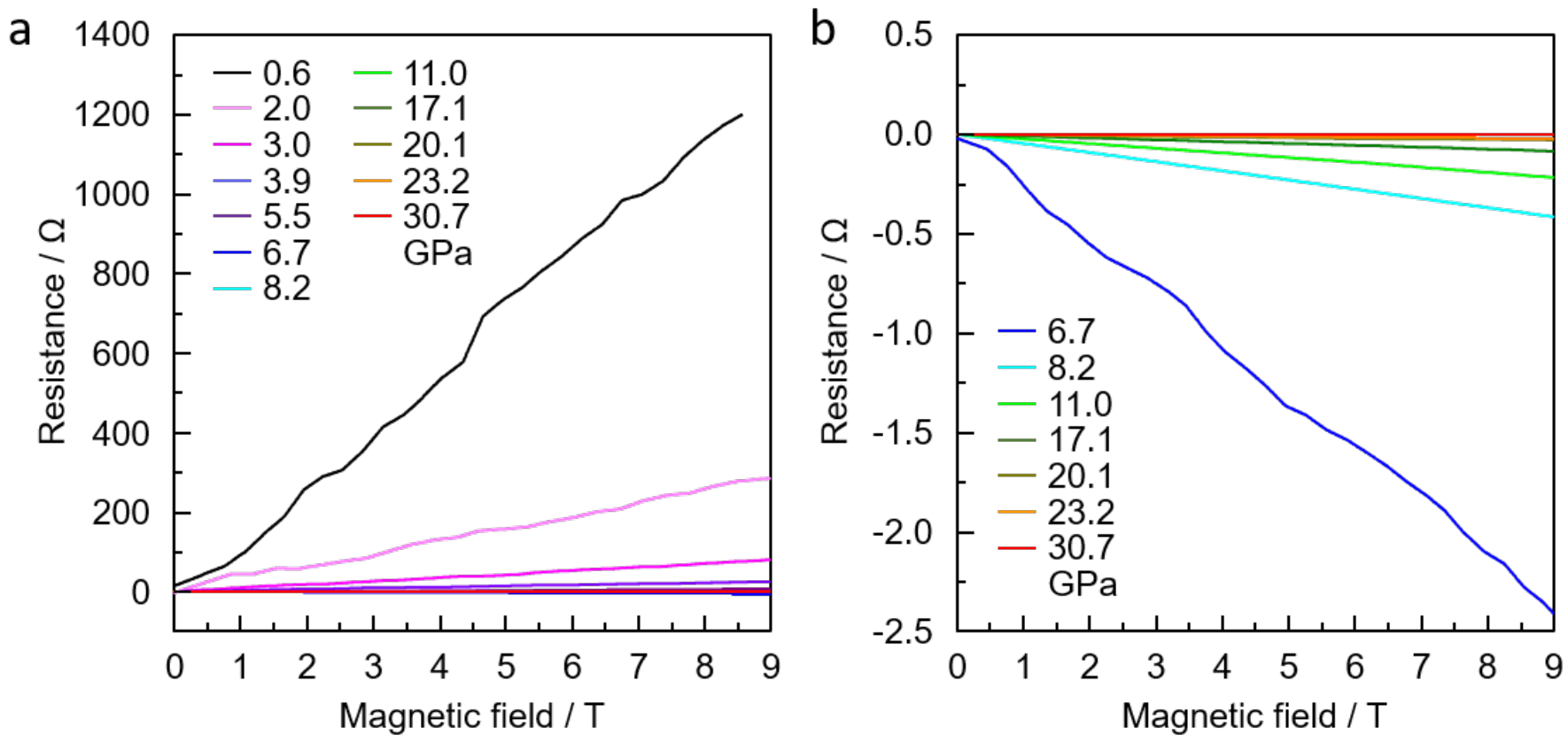

**Figure 2** (a) Hall resistance of $Mg_3Sb_2$ as a function of magnetic field under various pressures; (b) Enlarged Hall resistance of $Mg_3Sb_2$ in the vicinity of the carrier-type transition.

To explore the relationship between crystal structure and electronic transport properties under high pressure, we further employed *in situ* high-pressure Raman spectroscopy for $Mg_3Sb_2$. Three distinct peaks were observed at approximately 74.6, 119.3, and 151.5 $cm^{-1}$ at 0.1 GPa (**Figure 3a**). Based on our calculations (**Figure S7**) [49], these peaks are mainly attributed to the intra-layer vibrations of Sb and $Mg_2$ atoms, the inter-layer vibrations of Sb atoms, and the inter-layer vibrations of $Mg_1$ and $Mg_2$ atoms, respectively (**Figure 3b**). This assignment is consistent with previous reports. [26] Additionally, two relatively weak peaks at around 215.4 and 241.6 $cm^{-1}$ at 0.1 GPa are assigned to the vibrations of $Mg_2$ atoms.

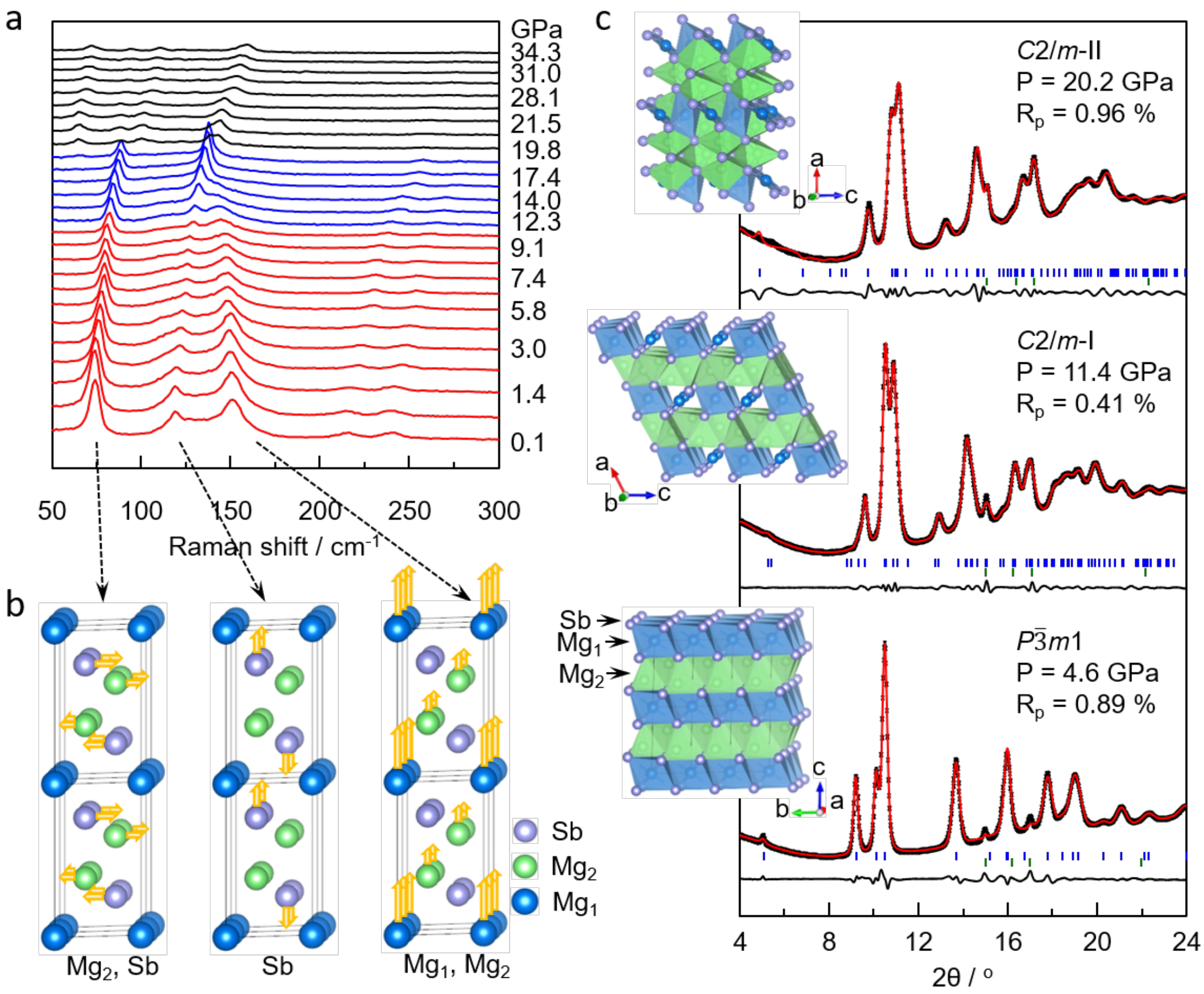

**Figure 3** (a) Raman spectra of $Mg_3Sb_2$ under various pressures; (b) Phonon mode symmetry and vibration directions of $Mg_3Sb_2$; (c) Diffraction profiles of high-pressure phases of $Mg_3Sb_2$ at 4.6 GPa, 11.4 GPa, and 20.2 GPa, respectively. Experimental and calculated patterns are indicated by black stars and red lines, respectively. Solid lines below each curve represent residual intensity. Vertical bars indicate peak positions of Bragg reflections for $Mg_3Sb_2$ in the $P\bar{3}m1$ space group at 4.6 GPa, $C2/m$-I at 11.4 GPa, and $C2/m$-II at 20.2 GPa (blue) and Re gasket (olive). The inset shows structures of $Mg_3Sb_2$ under ambient condition ($P\bar{3}m1$) and high pressure ($C2/m$-I and $C2/m$-II), respectively.

As pressure increases, the vibrational modes at 74.6 and 119.3 $cm^{-1}$ show a blue shift, while the mode at 151.5 $cm^{-1}$ tends to red shift below 9.1 GPa and further merges with the 119.3 $cm^{-1}$ mode above 12.3 GPa (**Figure 3a**). We analyzed these high-pressure Raman results in the context of structural evolution. It has been reported that octahedral $Mg_1$-Sb bonds in $Mg_3Sb_2$ are more compressible than tetrahedral $Mg_2$-Sb bonds under high pressure. [43] Following the structural phase transition, $Mg_1$ octahedra tilt into a square planar configuration due to the breaking of two opposing $Mg_1$-Sb bonds, while the $Mg_2$ tetrahedra distort into a layer with alternating tetrahedral and square pyramidal geometries (**Figure S8**). The red shift of the 151.5 $cm^{-1}$ mode is thus induced by the parallel movements of $Mg_1$ and $Mg_2$ along the *c* axis. Phonon frequencies are suppressed by the distortion of Mg units, which is accompanied by enhanced inter-layer interactions driven by pressure.

With further compression, the peak intensities of the 74.6 and 119.3 $cm^{-1}$ modes decrease sharply, while the peak signals around 215.4 and 241.6 $cm^{-1}$ almost vanish above 19.8 GPa (**Figure 3a**). Beyond this pressure, two new peaks appear at ~60 and 80 $cm^{-1}$. *In situ* high-pressure XRD experiments reveal that two peaks at around $2\theta$ = 6° and $2\theta$ = 10° merge above 18.2 GPa (**Figure S9**), which is consistent with a previous report [43]. The congruent high-pressure Raman and XRD results confirm an additional structural phase transition under higher pressure.

To confirm the high-pressure phase, we conducted structure searches for $Mg_3Sb_2$ at 15 GPa and 30 GPa. After 30 generations of evolution, we identified a predicted *C*2/*m* structure with a similar polyhedral stacking as the reported high-pressure structure [43] under higher pressure, and designated the reported and predicted two phases as *C*2/*m*-I and *C*2/*m*-II, respectively. (**Figure 3c**). The enthalpy difference of $Mg_3Sb_2$ structures relative to $P\bar{3}m1$ structure was calculated (**Figure S10)**. The $P\bar{3}m1$ structure has the lowest enthalpy at ambient pressure, while the *C*2/*m*-I structure becomes more stable above 5 GPa, consistent with high-pressure XRD and Raman results. With further compression, the *C*2/*m*-II structure exhibits lower enthalpy than the *C*2/*m*-I structure above 12.3 GPa, indicating higher thermodynamic stability under higher pressure. The deviation between the calculated transition pressure of 12.3 GPa and the experimental value of 20 GPa arises from the energy barrier induced by modified polyhedron stacking in these two $C2/m$ phases. Phonon spectra calculations demonstrate that $C2/m$-I remains dynamically stable up to 20 GPa and $C2/m$-II shows higher dynamical stability above this pressure (**Figure S11**). These results provide additional theoretical evidence for the pressure-induced structural phase transition in $Mg_3Sb_2$ under higher pressure.

Moreover, we calculated the electronic structures of $Mg_3Sb_2$ to reveal the experimental findings on transport properties during phase transitions. **Figure 4a** compares the calculated band structures of the $P\bar{3}m1$ phase at ambient pressure and the $C2/m$-I phase at 10 GPa. The high-symmetry points adopted in this work follow the definitions given in previous literature [50,51]. The $P\bar{3}m1$ phase exhibits an indirect band gap of ~0.56 eV with conduction band minimum (CBM) locating at the $U^*$ point, which is consistent with previous report [51]. To investigate the pressure-induced changes in the electronic structure, the electronic structure of the $P\bar{3}m1$ phase at 7 GPa is shown in **Figure S12**. Compared with the ambient-pressure case, the band gap at 7 GPa decreases by 0.05 eV. The valence band maximum (VBM) remains at the $\Gamma$ point, while the CBM shifts to the $K$ point. After the phase transition, the $C2/m$-I phase under 10 GPa has a small indirect band gap of 0.04 eV with VBM locating along the $X$-$\Gamma$ path and CBM locating along the $Y$-$X_1$ path, indicating a semiconductor-to-metal transition. The result indicates that the energy of CBM for the $C2/m$-I phase decreases significantly (~ 0.58 eV) compared with that of the $P\bar{3}m1$ phase. In semiconductors, the conduction type is governed by native defects and their formation energies. A defect with a lower formation energy is easier to form inside the lattice. The formation energy of donor defects decreases as the Fermi level decreases, while that of acceptor defects shows an opposite trend. So a low CBM energy (i.e., a low Fermi level) leads to a small donor formation energy, favoring n-type conduction [52,53]. This aligns with the carrier-type inversion above 6.7 GPa observed in our Hall effect measurements.

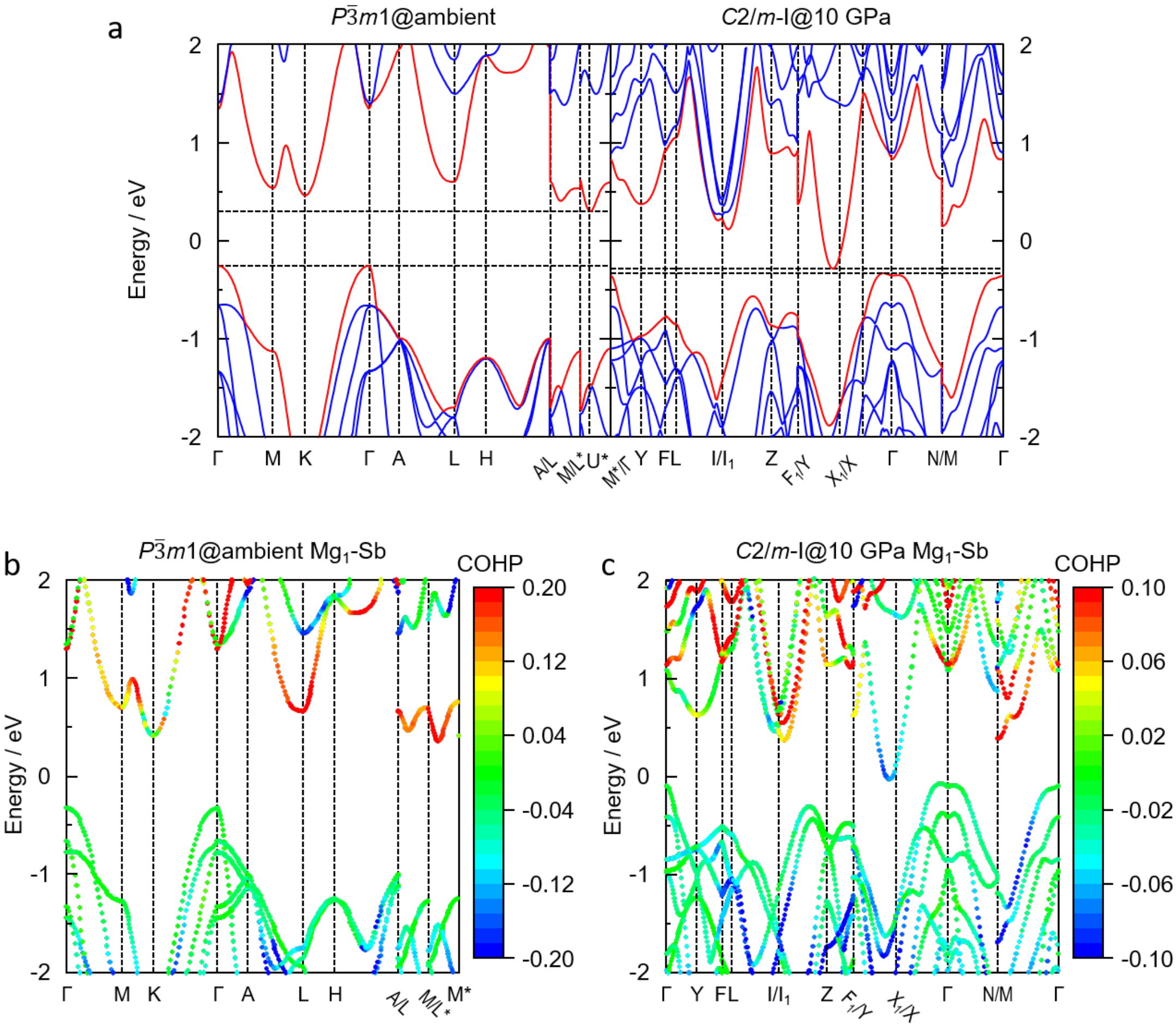

**Figure 4** Calculated band structures and band-resolved pCOHP. (a) Aligned bands of the $P\bar{3}m1$ phase at ambient pressure and the $C2/m$-I phase at 10 GPa with respect to the average Mg 1$s$ orbitals. Red lines indicate the band edges; (b) Band-resolved pCOHP of octahedral Mg-Sb for $P\bar{3}m1$ phase; (c) Band-resolved pCOHP of octahedral Mg-Sb for $C2/m$-I phase. The Fermi level is set as zero.

We further calculated the band-resolved projected Crystal Orbital Hamilton Population (pCOHP) for the $P\bar{3}m1$ and $C2/m$-I phases. Positive pCOHP values represent anti-bonding interactions, while negative values correspond to bonding interactions. As shown in **Figures 4b and 4c**, octahedral $Mg_1$ and Sb exhibit anti-bonding interactions at the CBM in the $P\bar{3}m1$ phase, but form bonding interactions in the $C2/m$-I phase. According to **Figures S13a and S13b**, the anti-bonding interactions between tetrahedral $Mg_2$ and Sb at the CBM of the $P\bar{3}m1$ structure translate to an almost non-bonding state at the CBM of the $C2/m$-I phase. Both of these changes in interactions contribute to the lowering of the CBM energy. Additionally, the $Mg_1$-$Mg_2$ bonding interactions at the CBM of the $P\overline{3}m1$ structure mentioned in the previous study [51] persist but with reduced strength (see **Figures S13c and S13d**).

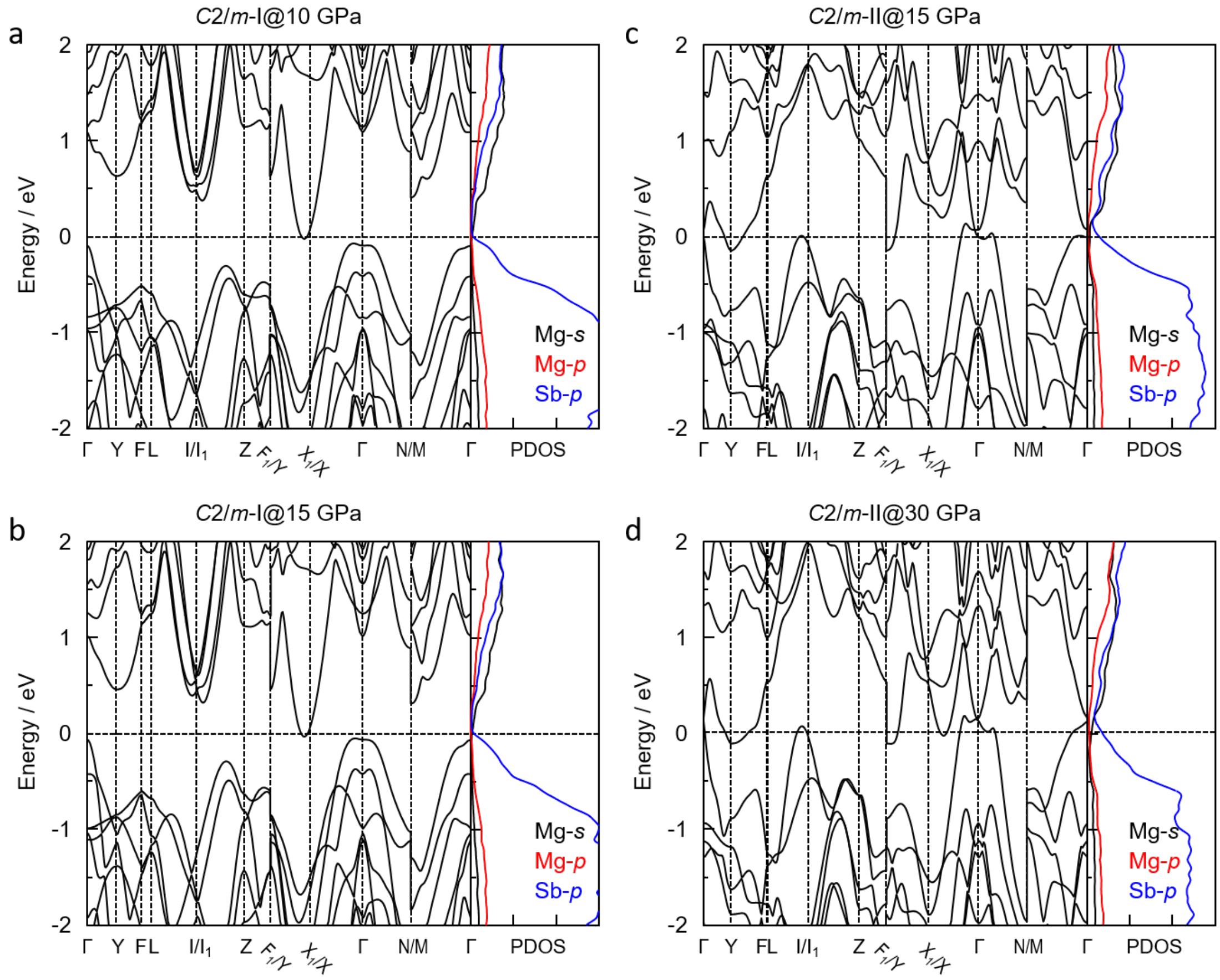

**Figure 5** (a)-(b) Band structures and PDOS of the *C*2/*m*-I phase at various pressures; (c)-(d) Band structures and PDOS of the *C*2/*m*-II phase at various pressures.

In addition, we calculated the band structures and partial density of states (PDOS) of both the *C*2/*m*-I and *C*2/*m*-II phases under various pressures. As shown in **Figures 5a and 5b**, the electronic structure of the *C*2/*m*-I phase undergoes only minor changes between 10 GPa and 15 GPa. For the *C*2/*m*-II phase (**Figures 5c and 5d**), the valence and conduction bands overlap, with additional bands crossing the Fermi level near the *Γ* point. PDOS results further indicate that the Sb-*p* electrons play a more prominent role at CBM of the *C*2/*m*-II phase compared with the case in the *C*2/*m*-I phase. Furthermore, the similarity in band characters between these two phases is consistent with the relatively constant $T_c$ value around 15 GPa, while the decrease in $T_c$ value above 20 GPa may be attributed to phonon hardening induced by high pressure. [54]

Based on the above results, we summarize the temperature-dependent phase diagram of $Mg_3Sb_2$ in **Figure 6**. As shown in **Figure 6a**, the resistivity $\rho$ at 300 K decreases exponentially near the semiconductor-to-metal transition pressure. Superconductivity emerges at ~ 9 GPa and $\rho$ at 300 K decreases gradually. For $T_c$, our replicated measurements demonstrate a dome-shaped relationship with pressure, with a maximum value of 3.3 K at 12.6 GPa. Combined with our *in situ* high-pressure XRD and Raman measurements, the superconducting transition is consistent with the pressure-induced structural phase transition to the *C*2/*m*-I phase.

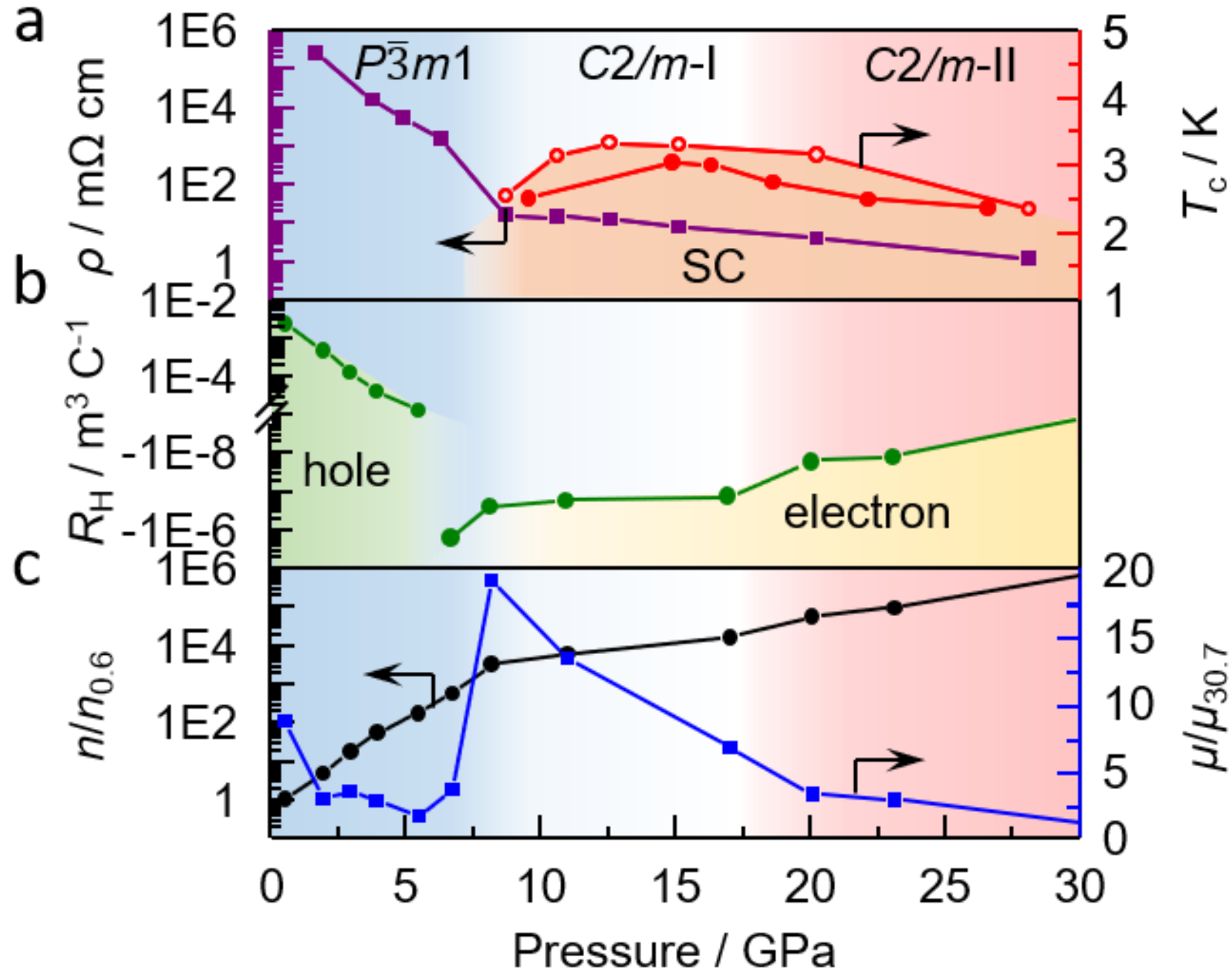

**Figure 6** (a) Pressure dependence of $T_c$ (red) and resistivity $\rho$ at 300 K (purple) for $Mg_3Sb_2$. Solid red dots are from run I and open red dots from run II; (b) Pressure dependence of the Hall coefficient (olive) for $Mg_3Sb_2$; (c) Pressure dependence of normalized carrier concentration $n$ (black) and carrier mobility $\mu$ (blue) for $Mg_3Sb_2$. $n_{0.6}$ denotes the carrier concentration measured at 0.6 GPa, and $\mu_{30.7}$ represents the carrier mobility obtained at 30.7 GPa.

**Figure 6b** presents the pressure-dependent Hall coefficient $R_H$ at 300 K. $R_H$ undergoes a dramatic shift from positive to negative, indicating a carrier-type transition from holes to electrons at ~ 7 GPa. Moreover, the absolute value of $R_H$ gradually approaches zero under higher pressure. We also extracted the carrier concentration and mobility at 300 K **(Figure 6c)**. Upon compression, carrier concentration continues to increase, but its growth rate slows down after ~ 7 GPa, which is consistent with the emergence of superconductivity. Carrier mobility increases significantly around 7 GPa, decreases with further pressure, and the downward trend moderated above 20 GPa. The transition pressures associated with carrier concentration and mobility at 300 K correspond to the structural transitions to the $C2/m$-I phase and $C2/m$-II phase, respectively. Therefore, we conclude that the pressure-induced structural phase transition from the $P\bar{3}m1$ phase to the $C2/m$-I phase triggers the semiconductor-to-superconductor transition at ~ 7 GPa. This transition is accompanied by a carrier-type transition from $p$-type to $n$-type. Our theoretical and experimental results further confirm an additional structural phase transition from the $C2/m$-I phase to the $C2/m$-II phase above ~ 20 GPa.

Compared with conventional strategies for optimizing thermoelectric materials under ambient conditions, high pressure offers substantially stronger tunability and can induce a wide range of emergent physical phenomena. From an electronic perspective, pressure-driven electronic evolution can optimize electrical transport properties and is therefore beneficial for enhancing thermoelectric performance. The structural phase transition reconstructs the $Mg_1$-Sb and $Mg_2$-Sb bonding states near the conduction band minimum (CBM), leading to an optimized band configuration. With increasing pressure, the substantial enhancement in electrical conductivity, accompanied by only a slight decrease in the Seebeck coefficient, can markedly improve the power factor[55-58]. In addition, the pressure-induced hole-to-electron carrier inversion is favorable for thermoelectric optimization, because $n$-type $Mg_3Sb_2$ intrinsically possesses superior transport properties[18,42]. In contrast, pressure-induced lattice distortions suppress the thermoelectric performance of $Mg_3Sb_2$. High pressure distorts the Mg-Sb polyhedra, reconstructs lattice vibrational modes, and strengthens interlayer interactions, ultimately leading to an increase in lattice thermal conductivity[59]. In particular, phonon hardening occurring above 20 GPa increases phonon frequencies and group velocities, thereby elevating the lattice thermal conductivity and weakening the thermoelectric performance[60]. Our work demonstrates that high pressure is an effective approach for identifying the key factors governing the thermoelectric properties of $Mg_3Sb_2$.

Moreover, high-pressure techniques not only provide valuable opportunities to optimize the thermoelectric parameters of existing thermoelectric materials, but also enable materials that are not thermoelectric under ambient conditions to be transformed into promising high-pressure thermoelectrics.

## CONCLUSIONS

In conclusion, we synthesized $Mg_3Sb_2$ single crystals and combined high-pressure measurements with structural searches and electronic structure calculations to investigate its high-pressure behavior. A semiconductor-to-metal transition was observed around 8.7 GPa, followed by a superconducting transition. This process is accompanied by a carrier-type transition from *p*-type to *n*-type. The pressure-dependent superconducting critical temperature $T_c$ exhibits a dome-shaped relationship, with a maximum value of 3.3 K at 12.6 GPa. High-pressure Raman and XRD measurements together with theoretical calculations confirm an additional structural transition above ~20 GPa. Our findings clarify the high-pressure phase diagram and electronic properties of $Mg_3Sb_2$, providing valuable insights for the study of TE materials under high pressure.

## ASSOCIATED CONTENT

The Supporting Information contains experimental section, high-pressure transport, XRD and calculation results of $Mg_3Sb_2$.

## AUTHOR INFORMATION

E-mail: qiyp@shanghaitech.edu.cn

jiongy@t.shu.edu.cn

chenguang_fu@zju.edu.cn

## ACKNOWLEDGMENT

This work was supported by the National Key R&D Program of China (Grant No. 2023YFA1607400) and the National Natural Science Foundation of China (Grant Nos. 52272265, 12474018). Chenguang Fu thanks the support from the National Natural

Science Foundation of China (Grant No. 52588301). Zhejiang Provincial Natural Science Foundation of China (LR25E020005). Jiong Yang thanks the support from the Key Research Project of Zhejiang Laboratory (No. 2021PE0AC02). The authors thank the support from the Analytical Instrumentation Center (# SPST-AIC10112914), SPST, ShanghaiTech University. The calculations were supported by the HPC platform of ShanghaiTech University. The authors thank the staff from BL15U1 at Shanghai Synchrotron Radiation Facility for assistance during data collection.

## REFERENCES

1 Snyder, G. J. & Toberer, E. S. Complex thermoelectric materials. *Nat. Mater.* **7**, 105-114 (2008).

2 He, J. & Tritt, T. M. Advances in thermoelectric materials research: Looking back and moving forward. *Science* **357**, eaak9997 (2017).

3 Chen, M. *et al.* Design of anomalous Nernst thermoelectric generators for giant power output. *Innovation (Camb)* **6**, 100995 (2025).

4 Lou, Q. *et al.* Recent advances and challenges in thermoelectrics toward near-room-temperature and high-temperature applications. *Chem. Soc. Rev.* **55**, 2265-2324 (2026).

5 Pei, Y., Wang, H. & Snyder, G. J. Band engineering of thermoelectric materials. *Adv. Mater.* **24**, 6125-6135 (2012).

6 Zhu, T. *et al.* Compromise and Synergy in High-Efficiency Thermoelectric Materials. *Adv. Mater.* **29** (2017).

7 Toriyama, M. Y. & Snyder, G. J. Topological insulators for thermoelectrics: A perspective from beneath the surface. *Innovation (Camb)* **6**, 100782 (2025).

8 Mao, H.-K., Chen, X.-J., Ding, Y., Li, B. & Wang, L. Solids, liquids, and gases under high pressure. *Reviews of Modern Physics* **90** (2018).

9 Pei, C. *et al.* Unveiling pressurized bulk superconductivity in a trilayer nickelate $Pr_4Ni_3O_{10}$ single crystal. *Sci. China-Phys. Mech. Astron.* **69**, 237011 (2025).

10 Pei, C. *et al.* Caging-Pnictogen-Induced superconductivity in skutterudites $IrX_3$ (X = As, P). *J. Am. Chem. Soc.* **144**, 6208-6214 (2022).

11 Pei, C. *et al.* Pressure induced superconductivity at 32 K in $MoB_2$. *Natl. Sci. Rev.* **10**, nwad034 (2023).

12 Pei, C. *et al.* Weakly anisotropic superconductivity of $Pr_4Ni_3O_{10}$ single crystals. *J. Am. Chem. Soc.* **148**, 1388-1396 (2026).

13 Zhu, S. *et al.* Pressure-induced superconductivity and topological quantum phase transitions in the topological semimetal $ZrTe_2$. *Adv. Sci.* **10**, 2301332 (2023).

14 Zhang, J. L. *et al.* Pressure-induced superconductivity in topological parent compound $Bi_2Te_3$. *Proc. Natl. Acad. Sci. USA* **108**, 24-28 (2010).

15 Gao, Y. *et al.* High thermoelectric performance of SnS under high pressure and high temperature. *Chinese Phys. B* **34**, 087201 (2025).

16 Jiang, Y. *et al.* Pressure-induced superconductivity and phase transition in PbSe and PbTe. *Chinese Phys. B* **33**, 126105 (2024).

17 Tamaki, H., Sato, H. K. & Kanno, T. Isotropic conduction network and defect chemistry in $Mg_{3+\delta}Sb_2$-based layered Zintl compounds with high thermoelectric performance. *Adv. Mater.* **28**, 10182-10187 (2016).

18 Zhang, J. *et al.* Discovery of high-performance low-cost *n*-type $Mg_3Sb_2$-based thermoelectric materials with multi-valley conduction bands. *Nat. Commun.* **8**, 13901 (2017).

19 Mao, J. *et al.* High thermoelectric cooling performance of *n*-type $Mg_3Bi_2$-based materials. *Science* **365**, 495 (2019).

20 Li, A., Fu, C., Zhao, X. & Zhu, T. High-Performance $Mg_3Sb_{2-x}Bi_x$ thermoelectrics: Progress and perspective. *Research (Wash D C)* **2020**, 1934848 (2020).

21 Li, A. *et al.* Demonstration of valley anisotropy utilized to enhance the thermoelectric power factor. *Nat. Commun.* **12**, 5408 (2021).

22 Li, A. *et al.* Chemical stability and degradation mechanism of $Mg_3Sb_{2-x}Bi_x$ thermoelectrics towards room-temperature applications. *Acta Mater.* **239**, 118301 (2022).

23 Li, A. *et al.* High performance magnesium-based plastic semiconductors for flexible thermoelectrics. *Nat. Commun.* **15**, 5108 (2024).

24 Shuai, J. *et al.* Recent progress and future challenges on thermoelectric Zintl materials. *Mater. Today Phys.* **1**, 74-95 (2017).

25 Ding, J. *et al.* Soft anharmonic phonons and ultralow thermal conductivity in $Mg_3(Sb,Bi)_2$ thermoelectrics *Sci. Adv.* **7**, eabg1449 (2021).

26 Peng, W., Petretto, G., Rignanese, G.-M., Hautier, G. & Zevalkink, A. An Unlikely Route to Low Lattice Thermal Conductivity: Small Atoms in a Simple Layered Structure. *Joule* **2**, 1879-1893 (2018).

27 Zhang, J., Song, L. & Iversen, B. B. Insights into the design of thermoelectric $Mg_3Sb_2$ and its analogs by combining theory and experiment. *npj Comput. Mater.* **5**, 76 (2019).

28 Imasato, K., Kang, S. D. & Snyder, G. J. Exceptional thermoelectric performance in $Mg_3Sb_{0.6}Bi_{1.4}$ for low-grade waste heat recovery. *Energy Environ. Sci.* **12**, 965-971 (2019).

29 Imasato, K., Kang, S. D., Ohno, S. & Snyder, G. J. Band engineering in $Mg_3Sb_2$ by alloying with $Mg_3Bi_2$ for enhanced thermoelectric performance. *Mater. Horiz.* **5**, 59-64 (2018).

30 Li, A. *et al.* Semiconductor-metal transition powers high-efficiency MgAgSb thermoelectrics. *Sci. Adv.* **11**, eadx7115 (2025).

31 Zhang, Z. *et al.* Ag rearrangement induced metal-insulator phase transition in thermoelectric MgAgSb. *Mater. Today Phys.* **25**, 100702 (2022).

32 Kanno, T. *et al.* Enhancement of average thermoelectric figure of merit by increasing the grain-size of $Mg_{3.2}Sb_{1.5}Bi_{0.49}Te_{0.01}$. *Appl. Phys. Lett.* **112**, 033903 (2018).

33 Hu, C., Xia, K., Fu, C., Zhao, X. & Zhu, T. Carrier grain boundary scattering in thermoelectric materials. *Energy Environ. Sci.* **15**, 1406-1422 (2022).

34 Shi, X. *et al.* Extraordinary n-type $Mg_3SbBi$ thermoelectrics enabled by yttrium doping. *Adv. Mater.* **31**, e1903387 (2019).

35 Gorai, P., Ortiz, B. R., Toberer, E. S. & Stevanović, V. Investigation of *n*-type doping strategies for $Mg_3Sb_2$. *J. Mater. Chem. A* **6**, 13806-13815 (2018).

36 Zhang, J., Song, L., Borup, K. A., Jørgensen, M. R. V. & Iversen, B. B. New insight on tuning electrical transport properties via chalcogen doping in *n*-type $Mg_3Sb_2$-based thermoelectric materials. *Adv. Energy Mater.* **8**, 1702776 (2018).

37 Nan, P. *et al.* Visualizing the Mg atoms in $Mg_3Sb_2$ thermoelectrics using advanced iDPC-STEM technique. *Mater. Today Phys.* **21**, 100524 (2021).

38 Ying, P. *et al.* Towards tellurium-free thermoelectric modules for power generation from low-grade heat. *Nat. Commun.* **12**, 1121 (2021).

39 Liu, Z. *et al.* Maximizing the performance of *n*-type $Mg_3Bi_2$ based materials for room-temperature power generation and thermoelectric cooling. *Nat. Commun.* **13**, 1120 (2022).

40 Ma, X. *et al.* Elevating thermoelectric performance in the sub-ambient temperature range for electronic refrigeration. *Innovation (Camb)* **6**, 100864 (2025).

41 Yang, H. *et al.* High-performance double-stage $Mg_3Bi_2$-based thermoelectric cooler. *The Innovation Mater.* **3**, 100130 (2025).

42 Wang, L., Li, A., Wu, X., Li, J. & Mori, T. Leveraging carrier mobility enables high-performance $Mg_3(Sb, Bi)_2$ thermoelectrics. *Natl Sci Rev* **13**, nwaf507 (2026).

43 Calderón-Cueva, M. *et al.* Anisotropic Structural Collapse of Mg3Sb2 and Mg3Bi2 at High Pressure. *Chem. Mater.* **33**, 567-573 (2021).

44 Dong, W. *et al.* Assessing structure of $Mg_3Bi_{2-x}Sb_x$ ($0 \le x \le 2$) at pressures below 40 GPa. *Journal of Materiomics* **10**, 837-844 (2024).

45 Woollam, J. A., Somoano, R. B. & O'Connor, P. Positive curvature of the $H_{c2}$-versus-$T_c$ boundaries in layered superconductors. *Phys. Rev. Lett.* **32**, 712-714 (1974).

46 Jones, C. K., Hulm, J. K. & Chandrasekhar, B. S. Upper critical field of solid solution alloys of the transition elements. *Rev. Mod. Phys.* **36**, 74-76 (1964).

47 Kimura, N., Ito, K., Aoki, H., Uji, S. & Terashima, T. Extremely high upper critical magnetic field of the noncentrosymmetric heavy fermion superconductor $CeRhSi_3$. *Phys. Rev. Lett.* **98**, 197001 (2007).

48 Hoshi, K., Kurihara, R., Goto, Y., Tokunaga, M. & Mizuguchi, Y. Extremely high upper critical field in $BiCh_2$-based (Ch: S and Se) layered superconductor $LaO_{0.5}F_{0.5}BiS_{2-x}Se_x$ (x = 0.22 and 0.69). *Sci. Rep.* **12**, 288 (2022).

49 Zhu, Y. *et al.* Violation of the $T^{-1}$ relationship in the lattice thermal conductivity of $Mg_3Sb_2$ with locally asymmetric vibrations. *Research (Wash D C)* **2020**, 4589786 (2020).

50 Setyawan, W. & Curtarolo, S. High-throughput electronic band structure calculations: Challenges and tools. *Comput. Mater. Sci.* **49**, 299-312 (2010).

51 Sun, X. *et al.* Achieving band convergence by tuning the bonding ionicity in *n*-type $Mg_3Sb_2$. *J. Comput. Chem.* **40**, 1693-1700 (2019).

52 Huo, H., Wang, Y., Xi, L., Yang, J. & Zhang, W. The variation of intrinsic defects in XTe (X = Ge, Sn, and Pb) induced by the energy positions of valence band maxima. *J. Mater. Chem. C* **9**, 5765-5770 (2021).

53 Xiong, Y. *et al.* High-throughput screening for thermoelectric semiconductors with desired conduction types by energy positions of band edges. *J. Am. Chem. Soc.* **144**, 8030-8037 (2022).

54 Dangić, Đ., Monacelli, L., Bianco, R., Mauri, F. & Errea, I. Large impact of phonon lineshapes on the superconductivity of solid hydrogen. *Commun. Phys.* **7**, 150 (2024).

55 Nabi, Z., Abbar, B., Mecabih, S., Khalfi, A. & Amrane, N. Pressure dependence of band gaps in PbS, PbSe and PbTe. *Comput. Mater. Sci* **18**, 127 (2000).

56 Li, C.-Y., Ruoff, A. L. & Spencer, C. W. Effect of pressure on the energy gap of $Bi_2Te_3$. *J. Appl. Phys.* **32**, 1733-1735 (1961).

57 Ovsyannikov, S. V. & Shchennikov, V. V. Pressure-tuned colossal improvement of thermoelectric efficiency of PbTe. *Appl. Phys. Lett.* **90**, 122103 (2007).

58 Ovsyannikov, S. V. *et al.* Giant improvement of thermoelectric power factor of $Bi_2Te_3$ under pressure. *J. Appl. Phys.* **104**, 053713 (2008).

59 Cai, J., Du, J., Pan, L., Chen, C. & Wang, Y. Effect of Te addition and texturing on the thermoelectric properties of TiS2 ceramics. *J. Appl. Phys.* **138**, 015701 (2025).

60 Li, S. *et al.* Anomalous thermal transport under high pressure in boron arsenide. *Nature* **612**, 459-464 (2022).

TOC

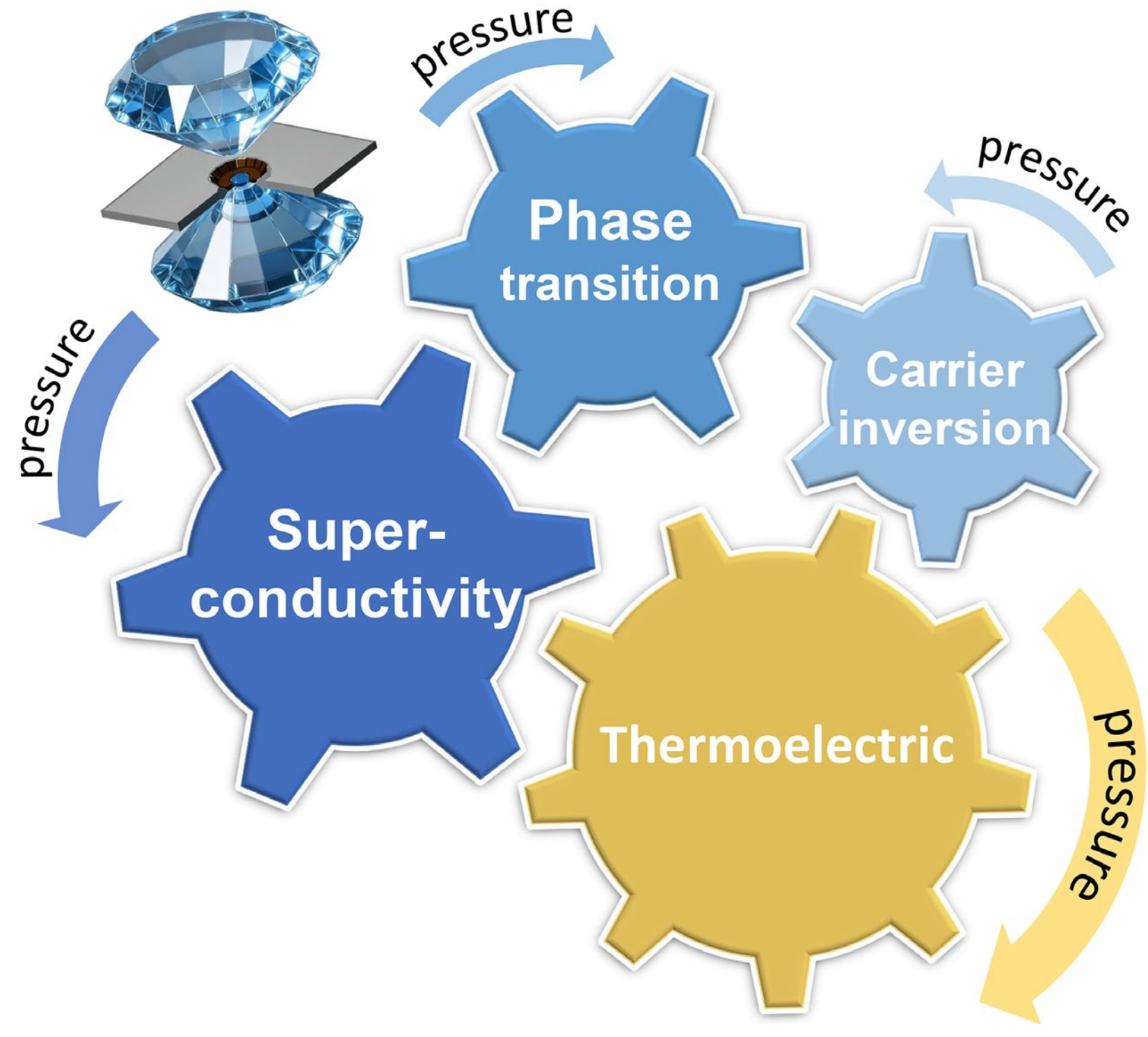
pressure
Phase
transition
pressure
Carrier
inversion
pressure
Super-
conductivity
Thermoelectric
pressure